\documentclass[12pt]{article}
\begin{document}
\newfam\msbfam
\batchmode\font\twelvemsb=msbm10 scaled\magstep1 \errorstopmode
\ifx\twelvemsb\nullfont\def\Bbb{\bf}
        \font\fourteenbbb=cmb10 at 14pt
	\font\eightbbb=cmb10 at 8pt
	\message{Blackboard bold not available. Replacing with boldface.}
\else   \catcode`\@=11
        \font\tenmsb=msbm10 \font\sevenmsb=msbm7 \font\fivemsb=msbm5
        \textfont\msbfam=\twelvemsb
        \scriptfont\msbfam=\tenmsb \scriptscriptfont\msbfam=\sevenmsb
        \def\Bbb{\relax\expandafter\Bbb@}
        \def\Bbb@#1{{\Bbb@@{#1}}}
        \def\Bbb@@#1{\fam\msbfam\relax#1}
        \catcode`\@=\active
	\font\fourteenbbb=msbm10 at 14pt
	\font\eightbbb=msbm8
\fi
\catcode`\@=11

\def\Z {{\Bbb Z}}
\def\R {{\Bbb R}}
\def\E {{\Bbb E}}
\def\scr{\cal}
\def\unit{\hbox to 3.3pt{\hskip1.3pt \vrule height 7pt width .4pt \hskip.7pt
\vrule height 7.85pt width .4pt \kern-2.4pt
\hrulefill \kern-3pt
\raise 4pt\hbox{\char'40}}}
\def\II{{\unit}}
\def\cM {{\cal M}}
\def\half{{\textstyle {1 \over 2}}}

\def    \beq    {\begin{equation}} \def \eeq    {\end{equation}}
\def    \bea    {\begin{eqnarray}} \def \eea    {\end{eqnarray}}
\def    \lf     {\left (} \def  \rt     {\right )}
\def    \a      {\alpha} \def   \lm     {\lambda}
\def    \D      {\Delta} \def   \r      {\rho}
\def    \th     {\theta} \def   \rg     {\sqrt{g}} \def \Slash  {\, /
\! \! \! \!}  \def      \comma  {\; , \; \;} \def       \pl
{\partial} \def         \del    {\nabla}
\newcommand{\mx}[4]{\left#1\begin{array}{#2}#3\end{array}\right#4}

\newcommand{\Dpp}{\Delta + \nu}
\newcommand{\Dmm}{\Delta - \nu}
\newcommand{\Dp}{\Delta_+}
\newcommand{\Dm}{\Delta_-}
\newcommand{\Ds}{\left(\Delta^2-\nu^2\right)}
\newcommand{\Pp}{\Pi_+}
\newcommand{\Pm}{\Pi_-}
\newcommand{\lp}{\ell_p}
\newcommand{\ie}{{\em i.e., }}
\newcommand{\eg}{{\em e.g. }}

\frenchspacing



\begin{titlepage}
\begin{flushleft}
       \hfill                      {\tt hep-th/0011282}\\
       \hfill                                  UG-00-21\\
       \hfill                                  G\"oteborg ITP preprint\\
      
\end{flushleft}
\vspace*{1.8mm}
\begin{center}
{\bf\LARGE Holographic Noncommutativity}\\

\vspace*{4mm}
{\large David S. Berman\footnote{berman@racah.phys.huji.ac.il}}\\
\vspace*{2mm}
{\small Racah Institute for Physics \\ 
 Hebrew University,  Jerusalem 91904, Israel}\\
\vspace*{3mm}
{\large Vanicson L. Campos,
Martin Cederwall, Ulf Gran, \\
Henric Larsson,  
Mikkel Nielsen, 
Bengt E.W. Nilsson\footnote{ vanicson, martin.cederwall, gran, 
solo, mikkel, tfebn@fy.chalmers.se}}\\
\vspace*{2mm}
{\small Institute of Theoretical  Physics\\
G\"{o}teborg University and Chalmers University of Technology\\
SE-412 96 G\"{o}teborg, Sweden}\\
\vspace*{3mm}
{\large and Per Sundell\footnote{p.sundell@phys.rug.nl}}\\
\vspace*{2mm}
{\small Institute for Physics, University of Groningen \\
\small Nijenborgh 4, 9747 AG Groningen, The Netherlands}\\

\vspace*{3mm}
\end{center}
\underbar{\bf Abstract:}
We examine noncommutative Yang--Mills and open string theories using
magnetically and electrically deformed supergravity duals. 
The duals are near horizon regions of D$p$-brane bound
state solutions which are obtained by using $O(p+1,p+1)$ 
transformations of D$p$-branes. The action of 
the T-duality group implies that the noncommutativity parameter
is constant along holographic RG-flows.
The moduli of the noncommutative theory, \ie the open string
metric and coupling constant, as well as
 the zero-force condition are shown to be 
invariant under the $O(p+1,p+1)$ transformation, \ie deformation independent.
 We find 
sufficient conditions, including zero force and constant dilaton in the 
$ISO(3,1)$-invariant D$3$ brane solution, for exact 
S-duality between noncommutative Yang--Mills and open string theories. 
These results are used to construct noncommutative field and string 
theories with ${\cal N}=1$ supersymmetry from the $T^{(1,1)}$ and 
Pilch--Warner solutions. The latter has a non-trivial zero-force condition
due to the warping.
\end{titlepage}


\section{Introduction}

String theory has provided a surprising connection between gauge
theories and supergravity. After taking certain decoupling limits one
can learn a great deal about the large $N$ limit of super-Yang--Mills by
studying its dual supergravity description \cite{malda,ofer}. This approach
of using the supergravity dual to obtain a description of the large N
limit of the theory has also been applied in many other
circumstances. In particular, it has been applied to the study of
noncommutative Yang--Mills theories \cite{maldaR,troels} as well as
field theories with fewer supersymmetries and RG-flows
\cite{RG,PW1,pilch,PW2,clif}.

A remarkable recent development has been the investigation of the
S-dual description of ${\scr N}=4$ noncommutative 
super-Yang--Mills (a D3 brane with background Neveu--Schwarz two-form
potential) which is provided by a noncommutative open string (NCOS)
theory \cite{gopa,lennat} on the three brane. In fact, one can
construct an NCOS theory on any D$p$-brane for $p \leq 5$. It is
possible to view these NCOS theories as ultraviolet completions of the
Yang--Mills theories with near critical background electric
fields (though for $p=5$ there is strong evidence that the theory
actually contains also a (closed) little string sector 
\cite{GMSS,AGM,harmark2}). 
Taking the strong coupling limit of the noncommutative open
string theory on the D4 brane leads to the so called OM theory which
in some sense acts as the M-theory for the open string theory
\cite{GMSS,us2,Kawano}. This is a five-brane in a background with
constant critical three-form potential, see also \cite{us1,kaw}.  
The supergravity
duals of these open string and membrane theories are known
\cite{townsend,Russo,martin,us3,eric,Harmark1,oz,Russo1,shiek}. One may 
then use these dual
supergravity solutions to determine many non-trivial properties of the
NCOS and OM theories
\cite{us3,Harmark1,Russo1,shiek}. In particular, the thermodynamics has been
investigated with non-extremal versions of these solutions
\cite{troels,Harmark1}.

In Section 2 we first review the decoupling limits that lead to
noncommutative theories on a D-brane and discuss the ultraviolet limit
of the near-horizon region of various supergravity backgrounds. 

In Section 3 we then use the supergravity dual picture to address the issue of
whether the noncommutativity parameter $\Theta$ can be tuned independently, or 
if a shift in $\Theta$ induces a variation of the other moduli. 
To examine this, we first give a simple form of
D$p$-brane bound states involving 
fluxes parametrized by an \mbox{$O(p+1,p+1)$} group element, as discussed in
\cite{Per}. We then go to the near-horizon limit and 
show how the decoupling limits can be obtained as
UV limits of the supergravity backgrounds. 
Using these general bound states we find that 
there is no deformation of the open string metric and coupling
constant, and that the probe brane potential is deformed by a 
multiplicative factor hence preserving the zero-force condition.
 Actually, the latter result will be crucial for 
the discussion of S-duality in Section 5.

We then exemplify these results in Section 4 using some simple 
models, namely the maximally symmetric extremal D$p$-branes, 
the $T^{(1,1)}$ extremal D$3$-brane and the D$1$--D$5$-brane system. 

In Section 5 we extend the deformation procedure to cases where 
one starts from the near-horizon solution (instead of the full 
brane solution). For D$3$ branes, 
we give a set of sufficient criteria for when the 
{\it critical electric} solution can be obtained from the magnetically
deformed near-horizon solution by type IIB S-duality (since 
the former cannot be directly obtained by an electric 
\mbox{$O(p+1,p+1)$} transformation). In particular, one condition is
that the configuration is restricted to the subspace where the 
probe brane potential vanishes. This is trivially satisfied for
maximally supersymmetric D$p$-branes. 

The  construction in Section 5 is useful since it does not require the full
brane solution, and in Section 6 we use this setup to construct an 
${\cal N}=1$ 
noncommutative Yang--Mills theory and its S-dual string theory based on
the Pilch--Warner solution \cite{pilch}. 
This theory has a non-trivial probe brane 
potential, and therefore illustrates the crucial importance for
S-duality played by the condition of vanishing potential.

\section{Open strings and supergravity duals\label{sec:os}}

The data governing the effective open string perturbation theory on a
D$p$-brane in a closed string background with string frame metric
$g_{MN}=g_{\mu\nu}\oplus g_{ij}$, dilaton $e^{\phi}$ and two-form
potential $B_{MN}$ is given by the open string two-point
functions \cite{sw}\footnote{Our definition of the noncommutativity parameter
$\Theta$ differs from the one in ref. \cite{sw} by a factor $2\pi$.
The present normalization yields exact equality between $\Theta$ and 
the deformation parameter $\theta$ in Section \ref{sec:nondef}.} 
together with the effective open string coupling.
 In this paper we use ten-dimensional spacetime indices
$M=0,\dots,9$, $(p+1)$-dimensional world volume indices
$\mu=0,\dots,p$ and $(9-p)$-dimensional transverse space indices
$i=p+1,\dots,9$. The two-point function is

\bea  <X^\mu(0) X^\nu(\tau)>&=&  -\alpha^\prime
G^{\mu\nu}\log{\tau} + i\pi\Theta^{\mu\nu}\epsilon(\tau)\ ,\nonumber\\ 
<X^i(0) X^j(\tau)> &=& -\alpha' g^{ij}\log {\tau}\ , \label{ncos2pt}\eea

where

\beq \label{ncos2} \alpha^\prime G^{\mu\nu} + \Theta^{\mu\nu} = 
\alpha^\prime \left( { 1 \over {g + 2 \pi \alpha^\prime
\langle{\cal F}\rangle}}\right) ^{\mu\nu} \ ,\label{gt} \eeq

and the effective open string coupling is

\beq \label{ncosg} G_{\rm O} = e^{\phi} \sqrt{ {\det(g + 2 \pi
\alpha^\prime \langle{\cal F}\rangle) \over  \det  g}}\ . \eeq

Here $\langle{\cal F}\rangle$ is the background value of the gauge
invariant field strength on the D$p$-brane:

\beq {\cal F}=dA+{1\over 2\pi \alpha'}(f^*B)\ ,\label{calf}
\eeq

that is

\beq \langle{\cal F}\rangle _{\mu\nu} = {1\over
2\pi\alpha'}(f^*B)_{\mu\nu}\ . \eeq

The symmetric part of (\ref{gt}), $G^{\mu\nu}$, is interpreted as the
open string cometric, its inverse being the open string metric. The
antisymmetric part $\Theta^{\mu\nu}$ is the noncommutativity
parameter. Thus the open string endpoints see an effective D-brane
world volume with metric

\beq ds^2(G)=G_{\mu\nu}dx^\mu dx^\nu\ , \label{osm}\eeq

and deformed algebra of functions with star-product based on the
Poisson structure

\beq \Theta=\Theta^{\mu\nu}\partial_\mu\partial^{'}_{\nu}\ .
\label{Theta} \eeq

We note the following useful identities:

\beq \label{osd1} G_{\mu\nu}=g_{\mu\nu}-(f^*B)_{\mu\rho}g^{\rho\sigma}
(f^*B)_{\sigma\nu}\ , \eeq \beq \label{osd2} G_{\rm
O}=e^{\phi}\left(\frac{\det G}{\det g}\right)^{1/4}\ ,\quad
\Theta^{\mu\nu}=-\alpha'g^{\mu\rho}(f^*B)_{\rho\sigma}G^{\sigma\nu}\
. \eeq

We are interested in examining a  D$p$-brane probe in a background
with large D$p$-brane charge. For large charge the background fields
are slowly varying functions of the distance between the probe brane
and the stack of source branes. The size of open
string fluctuations around the probe are governed by the effective
open string coupling $G_{\rm O}$ \cite{gopa} (while the size of closed
string fluctuations in the bulk are governed by the closed string
coupling $e^{\phi}$). In a region where
$G_{\rm O}<<1$ we have a one-parameter family of open string
quantum theories on the probe brane defined by the open string data
given above and two energy scales: the mass of the higgsed
$W$ bosons, above which the interactions between the probe and the
stack of source branes no longer is negligible, and the effective
Planck energy, which sets the cutoff for the closed string sector.

A fruitful feature of this setup is the possibility under certain
conditions to consider a near-horizon region of the background, which
defines a supergravity dual\footnote{To avoid confusion, we wish to
clarify that we shall distinguish between the full
brane solution and the near-horizon region and we shall always refer 
to the latter as the supergravity dual, even though this term could
of course also be used for the full brane solution. Note also that  
 the near-horizon limit in the NCOS case is defined in a slightly different 
manner from the standard one; this fact is discussed in more detail in Section
3.2. }, where the 
one-parameter family
extends all the way to the extreme UV. This amounts to taking the
separation (given in rescaled units in the near-horizon region)
between the probe brane and the source branes to infinity as to
decouple both $W$ bosons and closed strings while keeping a finite
(and small) open string coupling.  One important condition for the
UV-completion to be physically well-defined is that it should have a
stable ground state, \ie the zero-force condition should extend into
the UV. For finite separation it describes the flow from the
UV-completion down to the gauge theory in the extreme IR.

There are several important reasons why we prefer this holographic setup in
favour of simply scaling 'flat space' closed string moduli. Firstly,
it is not always the case that the total system becomes weakly coupled
so that it makes sense to describe the limits in a flat background. 
The advantage of the holographic setup is then of course that the 
non-perturbative aspects are automatically included in the description
of the limit. In particular, the relations between the critical scalings of the
various field strengths and tensions appear naturally in the bound state
solutions. Moreover, a bound state solution typically contains both 
magnetic {\it and} dual electric fluxes that couple to dual open branes. For
instance, as we shall see in Section 3, a magnetic NS flux typically 
comes with a 'crossed' electric RR flux and vice versa. Hence, instead of 
considering different source branes with electric or magnetic fluxes, one may
for any one given source brane 
obtain dual descriptions of the theory on the probe brane
simply by switching from one open string (\eg from an F1 to a D1) or brane
 description to another.
A third, perhaps less profound but nevertheless practical reason
is that due to the 'back-reaction' in the bound state solution from
switching on the noncommutative deformation parameter, the various issues
raised recently of whether and how the deformation acts on the other moduli
in the theory can be answered in a both precise and concrete fashion.

Of the various dual open branes\footnote{These include besides the
 open strings and OM cases also the OD$p$ cases in \cite{GMSS,harmark2}}
 referred to above, 
the two cases which are presently best understood are when the 
decoupled (and sometimes complete) UV-limits turn out to be a 
noncommutative gauge field theory (NCYM) or an open string 
theory (NCOS). A NCYM arises when the open string 
metric (\ref{osm}) diverges in units of ${\alpha}'$, while the 
NCOS arises in case this quantity
is fixed. The basic reason is that the rest-mass of an open string state
with oscillator number $N\geq 1$ is proportional to
$\sqrt{\vert G_{00}\vert (N-1)/\alpha'}$. Actually, the two possibilities are
sensitive to the signature of the noncommutativity parameter: NCYM
requires magnetic $\Theta^{\mu\nu}$ and NCOS electric
$\Theta^{\mu\nu}$. This relates to the fact that the asymptotic
geometry of the background is that of an array of smeared
F$1$-strings for NCOS and smeared D$(p-r)$-branes for NCYM with $r$
the rank of the tensor field generating the deformation.

Let us consider the case of a rank $2$ background two-form potential
$B_{\mu\nu}$ and let the limit, in which the separation between the
probe and the sources diverges, be controlled by some parameter
$\epsilon\rightarrow 0$. Suppose the closed string data in the brane
directions obey the following asymptotic scaling behaviour in the near-horizon 
region \cite{sw,liwu,chenwu} (for NCYM
$\alpha,\beta=0,\dots,p-2$ and $a,b=p-1,p$; for NCOS
$\alpha,\beta=0,1$ and $a,b=2,\dots,p$; $\lambda>0$):

\bea {\rm NCYM}&:& \qquad{g_{\alpha\beta}\over\alpha'}\sim
\eta_{\alpha\beta}\epsilon^{-\lambda}\ ,\qquad {B_{\alpha\beta}\over \alpha'}=0 \
,\nonumber\\ &&\qquad{g_{ab}\over \alpha'}\sim
\delta_{ab}\epsilon^{\lambda}\ ,\qquad {B_{ab}\over\alpha'}\sim
\epsilon_{ab}\epsilon^0\ ,\label{ncym}\\ &&\qquad e^{\phi}\sim
\epsilon^{{1\over 2}(5-p)\lambda}\ .  \nonumber\\&&\nonumber\\ {\rm
NCOS}&:&\qquad {g_{\alpha\beta}\over\alpha'}\sim
\eta_{\alpha\beta}\epsilon^{-\lambda}(1+U\epsilon^{\lambda}+\cdots)\
,\nonumber\\ &&\qquad{B_{\alpha\beta}\over
\alpha'}\sim\epsilon_{\alpha\beta}
\epsilon^{-\lambda}(1+V\epsilon^{\lambda}+\cdots)\ ,\qquad U+V\neq 0\
, \nonumber\\ && \qquad{g_{ab}\over \alpha'}\sim\delta_{ab}\epsilon^0\ ,\qquad
{B_{ab}\over\alpha'}=0\ ,\label{ncos}\\ &&\qquad e^{\phi}\sim
\epsilon^{-{1\over 2}\lambda}\ .\nonumber \eea

We remark that the physics depends on $\alpha'$ only via the tension
$g_{\mu\nu}/\alpha'$ and $B_{\mu\nu}/\alpha'$; thus the above limits
can be considered either as $\alpha'\rightarrow 0$ limits in a fixed,
flat background or as UV-limits in a supergravity dual \ie $r
\rightarrow \infty$, where the tension does not scale with $\alpha'$,
its scaling instead being determined by the energy scale set by the
distance between the probe and the source branes.

Both (\ref{ncym}) and (\ref{ncos}) yield infinite energies for massive
closed string states, whose rest-mass scales as
$\sqrt{\vert g_{00}\vert /\alpha'}$, while the energies of massive open string
states diverges for (\ref{ncym}) and remains finite for (\ref{ncos}),
as explained above. A more careful analysis \cite{oz2} based on calculations of
absorption cross-sections shows, however, that the massless closed
string sector only decouples for $p\leq 5$. In the case of $p=5$, \ie the
type IIB D$5$-brane, there is also compelling evidence that the theory also 
contains a (closed) little string sector \cite{harmark2}.

Next we are going to describe a construction of D$p$-brane bound states
with maximal rank deformations, and we shall then discuss the rank $2$
case using the above limits.

\section{Electric and magnetic deformations of open string data
\label{sec:nondef}}

We begin by describing how one may in a very efficient way 
construct the relevant deformed
background given an undeformed brane solution. We then explain how  
 the near-horizon limit of the
deformed solution is obtained. This is followed by the probing of the new
background solution via a D-brane probe. Finally, we analyse the open
string data given in the previous section on the probe D-brane in this
background.

\subsection{D$p$-brane bound states from $O(p+1,p+1)$ transformations}

Many examples of D$p$-brane bound states have been
constructed in the literature, see \eg
\cite{townsend,Russo,solutions,berg3}. The basic construction 
method is to combine
a series of diagonal T-dualities \cite{T-duality,berg}, constant
NS gauge transformations\footnote{There is an alternative method using
a boost/rotation between a compact and a non-compact direction
 instead of a gauge transformation. This method gives equivalent results 
to the method used in this paper and will not be discussed any further.}
and $SO(p,1)$ 
transformations, which together make up the T-duality group $O(p+1,p+1)$.
Recently, it was found that under a general transformation, the
RR potentials actually transform in a chiral $Spin(p+1,p+1)$ representation
\cite{fukuma,hassan}, and in \cite{Per} this was used to give a 
corresponding general parametrization of D$p$-brane bound states
as follows. Given a supergravity solution, one first T-dualizes 
in the directions where one wants to turn on NS fluxes, and then one
shifts $B_2$ with a constant in these directions. After this
one T-dualizes back again. In a more concise language, the
deformation  with parameter $\theta^{\mu\nu}/\alpha'$ is generated by
the following $O(p+1,p+1)$ T-duality group element 
\footnote{See \cite{fukuma,Per} for conventions and definitions of the 
various elements of $O(p+1,p+1)$ appearing in the following discussion.}

\beq \Lambda=\Lambda_0\dots
\Lambda_p\Lambda_{\theta/\alpha'}\Lambda_p\cdots \Lambda_0=
J\Lambda_{\theta/\alpha'}J=\Lambda_{-\theta/\alpha'}^T =
\mx{(}{ll}{1&0\\ \theta/\alpha'&1}{)}\ .  \eeq

Here $\theta^{\mu\nu}$ has dimension (length)$^2$ and carries indices
upstairs since it starts life on the T-dual world volume. In the 
NS-NS sector an element
$\left({a\atop c}{b\atop d}\right)$ 
transforms $E=g+B$ 
by means of a projective transformation, \ie $\tilde{E}=(aE+b)(cE+d)^{-1}$.
In the RR sector the anti-symmetric tensor fields can be shown to 
correspond to chiral spinors of the T-duality group. This follows from
mapping the inner product $i_{\mu}$ and one-form $dx^{\mu}$ onto the
annihilation and
creation operators that can be formed by taking linear combinations
of the relevant Dirac matrices. The result of this 
construction on the anti-symmetric tensor fields is given below. We shall
assume that the undeformed brane configuration (in the string
frame) is given by :

\bea ds^2&=&g_{\mu\nu}dx^\mu dx^\nu + g_{ij} dx^i dx^j\ ,\quad
B_2=\beta_2\ ,\nonumber\\
C&=&\omega dx^0\wedge\cdots
dx^p\wedge (1+\alpha)+\gamma\label{bc1} \eea

where $C$, and in particular $\gamma$, is a generalized sum of forms of different degree, and
$\beta_2$, $\gamma$ and $\alpha$ are transverse forms, \ie

\beq i_\mu\beta_2=i_\mu\alpha=i_\mu\gamma=0\ . \label{imu}\eeq

Here $i_{\mu}$ denotes inner product with the vector field associated
with $x^{\mu}$.  The deformed configuration is determined as
follows \cite{fukuma}:

\beq
\tilde{g}_{\mu\nu}+\tilde{B}_{\mu\nu}=\left[(g^{-1}+\theta/{\alpha}')^{-1}
\right]_{\mu\nu}\ ,\quad e^{\tilde{\phi}}=e^\phi\left(\frac{\det
\tilde{g}}{\det g}\right)^{1/4}\ ,\label{dcs} \eeq \beq \tilde{g}_{\mu
i}=\tilde{B}_{\mu i}=0\ ,\quad\tilde{g}_{ij}=g_{ij}\ ,\quad
\tilde{B}_{ij}=\beta_{ij}\ , \eeq \bea \tilde{C}&=&\omega
\left[\exp(-b_2) \exp\left(\frac1{2\alpha'}\theta^{\mu\nu}i_\mu
i_\nu\right)dx^0\wedge\cdots dx^p\right]\wedge(1+\alpha)\\
&&+\exp(-b_2)\wedge \gamma\ . \nonumber\eea

where $b_2$ is the deformed two-form in the brane directions. In
these directions we have

\beq d\tilde{s}^2_{p+1}=\left[g\left(1-(\theta
g/\alpha')^2\right)^{-1}\right]_{\mu\nu}  dx^\mu dx^\nu\ , \label{ds} \eeq
\beq b_2=-\frac1{2\alpha'} \left[g\theta
g\left(1-(\theta g/\alpha')^2\right)^{-1} \right]_{\mu\nu}
dx^\mu \wedge dx^\nu\ . \label{b} \eeq

In a frame where the metric and $(\theta g)^\mu{}_\nu$ are
block-diagonal we find that in the $k$'th $2\times 2$ block

\beq d\tilde{s}^2_{(k)}={ds^2_{(k)}\over h_k}\ , \quad
b^{(k)}_2=-{\theta_k\det{}_{(k)} g\over 2\alpha'h_k} \epsilon_{\mu\nu}
dx^\mu_{(k)}\wedge dx^\nu_{(k)}\ ,\quad
h_k=1+\left(\frac{\theta_k}{\alpha'}\right)^2\det{}_{(k)} g\ , \eeq

and that the dilaton is given by

\beq e^{2\tilde{\phi}}={e^{2\phi}\over \prod_k h_k}\ .  \eeq

We assume that the matrix $\delta^\mu_\nu- (\alpha')^{-2}
[(\theta g)^2]^\mu{}_\nu$ is invertible (for all $g_{\mu\nu}$), such that
the configuration in (\ref{ds}, \ref{b}) is non-singular. 
In the special case of a rank $2$ deformation we get

\beq d\tilde{s}^2={ds^{\prime 2}\over h}+ds_8^2\ ,\quad \tilde{B}_2=-
\frac{\theta\det{}' g}{2\alpha'h}\epsilon_{\mu\nu}dx^{\prime\mu}\wedge
dx^{\prime\nu}+\beta_2\ ,\label{r21} \eeq \beq
e^{2\tilde{\phi}}={e^{2\phi}\over h}\ , \quad h=1+\left({\theta\over
\alpha'}\right)^2\det{}' g\ ,\label{r22} \eeq \bea
\tilde{C}&=&\omega\left[{1\over h}dx^0\wedge\cdots dx^p+ {\theta\over
2\alpha'}\epsilon^{\mu\nu}i'_{\mu}i'_{\nu}dx^0\wedge\cdots \wedge
dx^p\right]\wedge(1+\alpha)\nonumber \\&& +\left[1+\frac{\theta\det{}'
g}{2\alpha'h}\epsilon_{\mu\nu}dx^{\prime\mu}\wedge
dx^{\prime\nu}\right]\wedge \gamma\ , \label{tildec} \eea
 
where $x^{\prime\mu}$ denote the two directions where the deformation
is non-trivial and $\det{}'$ is the $2\times 2$ determinant in this
space. 

It is worth mentioning that formally, the $O(p+1,p+1)$ transformations of the
solutions require the brane solution to be wrapped on a $(p+1)$-dimensional
torus, which may be decompactified after the transformation. However,
keeping compact directions, the D$p$-brane bound states give rise to
supergravity duals of the wrapped noncommutative open string theories of
\cite{Danielsson}.

\subsection{Near-horizon regions for deformed extremal branes \label{sec:nhl}}

For an extremal brane solution the metric in the brane directions
can be written as

\beq g_{\mu\nu}={\cal E}^2\eta_{\mu\nu}\ ,\label{vierbein}\eeq

where $\cal E$
 is a function of the transverse coordinates that
interpolates between the near-horizon region of the brane and an
asymptotical region such that 

\beq 0<{\cal E}<1 \ ,\label{extr}\eeq 

In the case of a rank $2$ magnetic deformation, that is $\det{}'g>0$, 
the near-horizon region is defined in the same way as the one 
of the undeformed configuration, \ie the supergravity dual of the 
(commutative) field theory on the brane. Thus, as we take $\alpha'
\rightarrow 0$ we keep fixed the following quantities:

\beq \mbox{Magnetic near-horizon:}\left.\begin{array}{c}\quad x^\mu,\
U^2\equiv {{\cal E}^2 \over \alpha'},\ \Phi^i\equiv{x^i\over \alpha'},\\[10pt]
\theta, \quad g^2_{YM}\equiv {g{(\alpha')}^{\frac{p-3}2}}\end{array}
\right\} \mbox{fixed}\ . \label{mnh} \eeq

Here it is important to note that
the canonical dimensions of the
fields are chosen such that if $ds^{2}/\alpha'$, $B_2/\alpha'$ and
$C_p/(\alpha')^{p/2}$ are held fixed in the $\alpha'\rightarrow 0$
limit, then the supergravity lagrangian is finite.
In the absence of other dimensionful parameters, in the
$\alpha'\rightarrow 0$ limit it is $U$ as a function of the vacuum
expectation values of the scalar fields $\Phi^i$, that sets the energy
scale on the probe brane (it is assumed that $x^i$ are coordinates of
dimension length). In the UV limit, $U\rightarrow \infty$, one
recognizes the NCYM limit (\ref{ncym}) with the undeformed 
dilaton scaling as

\beq  e^\phi \sim g_{\rm YM}^2 U^{p-3}\ ,\label{cond} \eeq

where $g_{\rm YM}^2$ is the Yang--Mills coupling in $p+1$ dimensions.

Note that the requirement that the scale is determined by $U$
implies that the pull-backs of the transverse background components
scales to zero in the UV. For example, the scalar kinetic term will
scale as

\beq f^*(g_{ij}dx^i dx^j) \sim \alpha' U^{-2}dx^\mu dx^\nu
\partial_\mu \Phi^i \partial_\nu \Phi^j \bar{g}_{ij}\ ,\label{uv1} \eeq

where $f$ denotes the embedding of the probe brane in spacetime. Here 
$\bar{g}_{ij}$ is a dimensionless function of transverse
coordinates which does not scale in the UV and the scalar gradients
$\partial_\mu \Phi^i$ are fixed in the UV.  

We remark that in many cases the field theory near-horizon geometry
is warped \cite{Pope}, such that the near-horizon geometry
factorizes into that of a $(p+2)$-dimensional anti-de Sitter 
space and an $(8-p)$-dimensional internal space in such a way that
the energy scale $U$ 
defined in (\ref{mnh}) and the energy scale $u$ of the near 
horizon anti-de Sitter part of the spacetime metric are related as follows:

\beq
ds^2=\Omega^2 ds^2_{AdS_{p+2}}+d\hat{s}^2_{8-p}\ ,\quad
ds^2_{AdS_{p+2}}=u^2 dx_{p+1}^2+{du^2\over u^2}\ ,\eeq
\beq
U^2= \Omega^2 u^2\ ,\label{wf}
\eeq

where the warp factor $\Omega$ is a dimensionless (non-constant) 
function of the internal coordinates in the line element $d\hat{s}_{8-p}^2$. 
This is equivalent
to the internal metric defined in (\ref{uv1}) being
block-diagonal as follows: 

\beq
d\bar{s}^2_{9-p} = \Omega^4 (du^2+u^2 d\hat{s}_{8-p}^2)\ .\eeq

As we shall see in the next subsection, this gives rise to a probe 
brane potential $V$. Excitations that break the zero-force condition have a 
potential energy that scales like $V\sim
g_{\rm YM}^{-2}U^{4}\rightarrow \infty$ in the UV, so they will be 
frozen out in taking the UV limit.
In what follows it will be useful to define the map $i$ as follows

\beq i:V^{-1}(0)\mapsto M^{10}\ ,\label{locemb} \eeq

it is the embedding of the submanifold where the potential energy of a
probe D-brane vanishes in the ten-dimensional spacetime
$M^{10}$. This submanifold is called {\it{the vanishing locus}}. 

In the case of an electric deformation we have $\det{}' g<0$, so the
`field theory' near-horizon limit (\ref{mnh}) would yield a
supergravity dual with a singularity at a finite energy scale (where
$h$ would vanish). 
To avoid this, the electric near-horizon region is instead defined by the
following critical 
limit\footnote{$g$ is here the closed string coupling constant.}, 
as $\alpha' \rightarrow 0$:
\beq \mbox{Electric near-horizon:}\quad {x^M\over \sqrt{\alpha'}}
,\ {\cal E},\ {g},\ {\omega}\ \mbox{fixed}\ ;
\quad {\theta\over \alpha'}\rightarrow 1\ .\label{enh} \eeq

Provided that there are no other dimensionful parameters than $\theta$
and $\alpha'$, then, since all coordinates are kept fixed in units of
$\alpha'$, the electric near-horizon region contains the original asymptotic
region. The UV limit now corresponds
to $h=1-{\cal E}^4\rightarrow 0$, which can be seen to 
reproduce the NCOS limit
(\ref{ncos}) (using (\ref{extr})). We remark that the
spacetime metric approaches that of a smeared string in the UV; taking
the electric deformation in the $0$ and $1$ directions we have

\beq {\rm NCOS}:\quad {ds^2\over
\alpha'}~\sim\!\!\!\!\!\!\!{}^{{}^{\scriptscriptstyle{\rm UV}}}~
h^{-1}(-dx_0^2+dx_1^2)+dx^2_2+...+dx_p^2+\bar{g}_{ij}dx^i dx^j\ , \label{smeared} \eeq

where $\bar{g}_{ij}$ is defined in (\ref{uv1}).
It follows that all closed string states except the massless states
with momentum along the string are frozen out from the perturbative
spectrum in the sense that their energy diverges. Similarly, from
(\ref{uv1}) it follows that for the NCYM limit the asymptotic metric
on the vanishing locus approaches that of a smeared D$(p-2)$-brane:

\beq {\rm NCYM}:\quad {ds^2\over
\alpha'}~\sim\!\!\!\!\!\!\!{}^{{}^{\scriptscriptstyle{\rm UV}}}~
u^2(dx^2_0+...+dx^2_{p-2})+u^{-2}(dx^2_{p-1}+dx^2_p+
i^*d\Phi^2_{p+1}+...+i^*d\Phi^2_{9})\ . \eeq

and the perturbative closed string spectrum reduces in the UV to the
massless modes with momentum parallel to the smeared brane.  Hence, at
the perturbative level the dynamics of the D$p$-brane is decoupled
from the bulk. 

The deformed open string data in the near-horizon region is found by
inserting the deformed closed string configuration (\ref{dcs}) into
(\ref{osd1}) and (\ref{osd2}). For the open string metric and
noncommutativity parameter we find

\beq \tilde{G}_{\mu\nu}=g_{\mu\nu}\ ,\quad
\tilde{\Theta}^{\mu\nu}=\theta^{\mu\nu}\ .  \eeq

Hence the open string metric is undeformed. Using (\ref{ncosg}) we
then find that also the  open string coupling is undeformed:

\beq \tilde{G}_{\rm O}=e^{\tilde{\phi}}\left(\frac{\det
\tilde{G}}{\det \tilde{g}}\right)^{1/4}
=e^\phi\left(\frac{\det \tilde{g}}{\det g}\right)^{1/4}
\left(\frac{\det g}{\det \tilde{g}}\right)^{1/4}=e^\phi \ .  \eeq

Actually, starting from a more general brane configuration in which
there are already NS fluxes inside the world volume one still
finds that

\beq \tilde{G}_{\mu\nu}=G_{\mu\nu}\ ,\quad
\tilde{\Theta}^{\mu\nu}=\Theta^{\mu\nu}+\theta^{\mu\nu}\ , \quad
\tilde{G}_{\rm O}=G_{\rm O}\ . \eeq

We also see that the noncommutativity parameter must remain constant
along the RG-flow\footnote{This was observed using the explicit supergravity
solution in \cite{shiek}, and deduced from a non-renormalization argument
using supersymmetry already in \cite{intriligator}.}. 
One might expect the noncommutativity to vanish in
the IR because the Neveu--Schwarz two-form vanishes. This is however not
the case due to the compensating scaling of the metric. A further
understanding of this would be desirable, even though it seems
 that the effects of the constant noncommutativity parameter 
vanishes in the IR, since fields become slowly varying in this limit. 

 We remark that for an NCOS theory, it follows from (\ref{enh}) that in the UV
the open string tension and noncommutativity scale as

\beq {ds^2(\tilde{G})\over \alpha'}\sim {\eta_{\mu\nu}dx^\mu
dx^\nu\over \alpha'}\ ,\quad \tilde{\Theta}\sim
\alpha'\epsilon^{\alpha\beta}\partial_\alpha{\partial}^{'}_\beta\ , 
\label{osd5}\eeq

such that indeed the inverse of the effective open string tension
is equal to the noncommutativity parameter (after going to fixed
coordinates, the coordinates in (\ref{osd5}) obey (\ref{enh})).

\subsection{Non-deformation of zero-force condition\label{sec:zfc}}

It was shown in \cite{Per} that the zero-force condition for the probe brane
in the supergravity dual is not deformed by a noncommutative deformation. 
To examine this, one uses the UV expansion of the
Born--Infeld lagrangian, which is discussed in Appendix \ref{app:2} (see
(\ref{dbiexp})), and that the
contribution to the potential from the WZ term is given by

\bea \tilde{\cal L}_{\rm WZ} &=& \omega dx^0\wedge\cdots \wedge dx^p
\wedge \exp\left(\frac{\theta^{\mu\nu}}{2\alpha'}i_\mu i_\nu
\right)\exp\left(\frac12 F_{\mu\nu}dx^\mu\wedge dx^\nu\right)
\nonumber\\ &=&\omega\sqrt{\det(1+\theta F)} dx^0\wedge\cdots \wedge
dx^p\ ,\label{vev} \eea

where the bound state solution (\ref{r21}) and (\ref{tildec}) have been used. 
Note that the determinant is a square of
Pfaffians, so that the last expression is indeed a finite polynomial
in traces of $\theta F$. The deformed potential is therefore given by

\beq \tilde{V}=\sqrt{\det(1+\theta F)}~V\ ,\label{tilv} \eeq

where

\beq
V=(\alpha')^{-\frac{p+1}{2}}\left(e^{-\phi}\sqrt{-{\rm det}g_{\mu\nu}}-
\omega\right) \label{pot}\eeq

is the undeformed potential, which shows that the zero-force
condition is not deformed by turning on a noncommutativity parameter.

\section{Three examples}

In this section we give three examples where the warp factor $\Omega$
in (\ref{wf}) is constant, and the probe brane potential is 
identically zero. A case with non-trivial warp factor will be examined in 
Section 6.

\subsection{Stack of maximally symmetric extremal D$p$-branes\label{sec:ex1}}

As the canonical example we consider a stack of maximally
symmetric extremal D$p$-branes ($p\leq 6$):

\bea ds^2&=&H^{-{1\over 2}}dx^2+H^{1\over 2}(dr^2+r^2d\Omega^2)\
,\nonumber\\ e^{2\phi}&=&g^2H^{\frac{3-p}2}\ ,\qquad
H=1+{gN(\alpha')^{\frac{7-p}2}\over r^{7-p}}\ ,\label{dbrane} \\
C&=&\frac1{gH}dx^0\wedge \cdots\wedge
dx^p+(7-p)N(\alpha')^{\frac{7-p}2} \epsilon_{7-p}\ , \nonumber \eea

where $d\epsilon_{7-p}$ is the volume element on the transverse
$(8-p)$-sphere. After a rank $2$ deformation, using the T-duality
transformation described above, the magnetically deformed configuration is

\bea 
d\tilde s^2&=&H^{-{1\over 2}}\left(-dx_0^2+\ldots+dx_{p-2}^2
+{1\over h}(dx^2_{p-1}+dx_p^2)\right)+H^{1\over 2}(dr^2+r^2d\Omega^2)\
,\nonumber\\ e^{2\tilde\phi}&=&{1\over h}g^2H^{\frac{3-p}2}\ , \nonumber \\
\tilde B&=&-{\theta^{p-1,p}H^{-1}\over\alpha'h}dx^{p-1}\wedge dx^p \ ,\label{Mdbrane}
  \\
\tilde C&=&\frac1{gHh}dx^0\wedge \cdots\wedge
dx^p+(7-p)N(\alpha')^{\frac{7-p}2}(1-\tilde B)\wedge \epsilon_{7-p}\nonumber \\
&&-{\theta^{p-1,p}\over gH\alpha'}dx^0\wedge\ldots\wedge dx^{p-2}\ , 
\nonumber\\
\mbox{where}&& h\equiv1+\left({\theta^{p-1,p}\over\alpha'}\right)^2H^{-1}\ , 
\nonumber
\eea

and the electrically deformed one is

\bea 
d\tilde s^2&=&H^{-{1\over 2}}\left({1\over h}(-dx_0^2+dx_1^2)+dx_2^2
+\ldots+dx_p^2\right)+H^{1\over 2}(dr^2+r^2d\Omega^2)\
,\nonumber\\ e^{2\tilde\phi}&=&{1\over h}g^2H^{\frac{3-p}2}\ , \nonumber \\
\tilde B&=&{\theta^{01}H^{-1}\over\alpha'h}dx^0\wedge dx^1 \ ,\label{Edbrane}
  \\
\tilde C&=&\frac1{gHh}dx^0\wedge \cdots\wedge
dx^p+(7-p)N(\alpha')^{\frac{7-p}2}(1-\tilde B)\wedge \epsilon_{7-p}\nonumber \\
&&-{\theta^{01}\over gH\alpha'}dx^2\wedge\ldots\wedge dx^p\ ,
\nonumber\\
\mbox{where}&&
h\equiv1-\left({\theta^{01}\over\alpha'}\right)^2H^{-1}\ .\nonumber
\eea

We define the magnetic and electric
supergravity duals by the resulting near-horizon limits, taking
$\alpha' \rightarrow 0$,

\bea \mbox{Magnetic near-horizon}&:& \quad x^\mu,\ \frac{r}{\alpha'},\ 
\theta,\ g_{\rm YM}^2\equiv g(\alpha')^{\frac{p-3}2}\quad
\mbox{fixed}\ .\label{ymnh}\\
\mbox{Electric near-horizon}&:& \quad {x^\mu\over \sqrt{\alpha'}},\
{r\over \sqrt{\alpha'}},\ g\quad \mbox{fixed}\ ;\quad
{\theta\over\alpha'}\rightarrow 1\ .\label{osnh}\eea

These limits are easily seen to arise from the brane solutions by first 
introducing the relevant rescaled radial coordinate in the two, \ie
electric and magnetic, cases. The other definitions in (\ref{ymnh}) and (\ref{osnh}) then 
follow by demanding that the quotients of the 
background fields and powers of $\alpha'$ defined in Section 3.1 should be
independent of $\alpha'$. 
The near-horizon geometry thus obtained interpolates between conformal
$AdS_{p+2}\times S^{8-p}$ and an array of smeared D$(p-2)$-branes for
NCYM and F$1$-strings for NCOS. We verify (\ref{ncym}) and (\ref{ncos})
as follows:

\bea \mbox{NCYM}&:& \epsilon=H\ ,\lambda={1\over 2}\ .\label{ncymuv}\\
\mbox{NCOS}&:& \epsilon=h\ ,\lambda=1\ ,U=-{1\over 2}\ ,V=-1\ .
\label{ncosuv}\eea

Note that for NCYM it follows from (\ref{ymnh}) that $(\alpha')^2 H$
is fixed, \ie independent of $\alpha'$,
 in the near-horizon region, where it vanishes in the UV,
so that strictly speaking we should take $\epsilon=(\alpha')^2 H/\ell^2$
in (\ref{ncymuv}) for some fixed length scale $\ell$. We also remark that
(\ref{cond}) is indeed obeyed with the definition of
the Yang--Mills coupling made in (\ref{ymnh}). 
Using (\ref{osd1}) and (\ref{osd2}) we
compute the open string data in the complete brane configuration as
follows:

\beq \tilde{G}_{\mu\nu}=H^{-{1\over 2}}\eta_{\mu\nu}\ ,\quad
\tilde{G}_{\rm O}=g H^{\frac{3-p}4}\ , \quad
\tilde{\Theta}^{\mu\nu}=\theta^{\mu\nu}\ ,\eeq

where $\theta^{\mu\nu}$ is magnetic or electric as the case may
be. From (\ref{ymnh}) and (\ref{osnh}) it follows that in the near
horizon region the following open string data are fixed:

\beq {ds^2(\tilde{G})\over \alpha'}= {H^{-{1\over 2}}dx^2\over
\alpha'} \ ,\quad\tilde{G}_{\rm O}= g H^{\frac{3-p}4}\ , \quad
\tilde{\Theta}=\theta^{\mu\nu}\partial_\mu\partial^{'}_\nu\ . \eeq

For NCOS, it follows from (\ref{osnh}) that the above quantities are 
indeed finite in the UV limit (\ref{ncosuv}). By going to fixed 
coordinates, which requires the introduction of some arbitrary length scale 
$\sqrt{\alpha'_{\rm eff}}$, it follows that
the inverse of the 
open string tension and the spatio-temporal noncommutativity parameter
are indeed equal and given by $\alpha'_{\rm eff}$ in the UV. 

For NCYM, it follows from (\ref{ymnh}) that in the UV limit (\ref{ncymuv}) 
the noncommutativity $\tilde{\Theta}$ is finite, while the open
string tension $ds^2(\tilde{G})/\alpha'$ diverges.
However, defining a fixed metric $ds^2(\bar{G})$ and length scale
$\ell$ by

\beq \mbox{NCYM}\ :\ {ds^2(\tilde{G})\over \alpha'}\sim
\epsilon^{-\frac12} {ds^2(\bar{G})\over \ell^2}\ ;\quad ds^2(\bar{G})\
, \ell\quad \mbox{fixed}\ ,\label{barG}\eeq

yields a finite kinetic term in the Born--Infeld lagrangian of the
 probe D-brane:

\bea S_{\rm DBI}&=&-\frac{1}{4g_{\rm YM}^2}\int d^{p+1}x
\sqrt{-\det{\bar{G}}}\bar{G}^{\mu\rho}\bar{G}^{\nu\sigma}F_{\mu\rho}
F_{\nu\sigma}+\cdots\ ,\label{dbil} \\
g_{YM}^{2}&=&\epsilon^{\frac{p-3}4}\ell ^{p-3}\tilde{G}_{\rm O}\ ,
\label{gYM}\eea

where the fixed Yang--Mills coupling $g_{\rm YM}$ can be seen to agree
with the definition (\ref{ymnh}). The ellipses in (\ref{dbil})
represent more terms that are finite in the UV, while the leading
divergent terms cancel against the WZ-term due to the zero-force
condition, as explained in Section \ref{sec:nhl} and
\ref{sec:zfc}. From the discussion in Appendix \ref{app:2}, it follows
that the complete probe brane lagrangian, including the WZ-term, can
be written as the noncommutative lagrangian (\ref{ncs}) in the extreme 
UV region, where we can take

\beq \bar{G}_{\mu\nu}=\eta_{\mu\nu}\ ,\quad \bar{g}_{ij}=\delta_{ij}\
.\eeq

As discussed in Section \ref{sec:os}, it has important consequences 
for the interpretation of the resulting
NCYM, that the open string coupling
diverges in the UV for $p>3$ and vanishes for $p=2$, as follows from
(\ref{gYM}). For $p=3$, 
the NCYM is a UV complete theory since the open string coupling 
is fixed in the UV. For $p=4$, 
the noncommutative lagrangian (\ref{ncs}) should be thought of as an
effective field theory description valid at energies below
$g_{\rm YM}^{-2}$. The bound state has a critical electric RR three-form.
The UV completion is therefore an open D$2$-brane theory in five dimensions.
By lifting this theory to M-theory along a magnetic circle of radius 
$g_{\rm YM}^2$ one finds that it is dual to the noncommutative 
M-theory five-brane \cite{GMSS,us2}. The five brane KK modes appears
in the D$4$-brane NCYM theory as noncommutative instantons with
energy $g_{\rm YM}^{-2}$. For $p=5$ the bound state has a critical electric
RR four-form, and the UV completion involves an open D3-brane.

\subsection{The $T^{(1,1)}$ solution}

We next consider an extremal D$3$-brane with less supersymmetry.
 The $T^{(1,1)}$ {$\scr N$}=1 brane solution is \cite{duff1}:

\bea ds^2&=&H^{-{1\over 2}}dx^2+H^{1\over 2}(dr^2+r^2d\Omega^2_{T})\
,\nonumber\\ e^{2\phi}&=&g^2\ ,\qquad H=1+{gN(\alpha')^{2}\over
r^{4}}\ .\label{T11} \eea

Here $d\Omega_{T}$ is the line element of $T^{(1,1)}$, which implies that
 the near-horizon region is $AdS_{5}\times T^{(1,1)}$. This near-horizon 
region is the AdS/CFT dual of an {$\scr N$}=1 super-conformal field theory 
\cite{wk1}. Comparing this metric with (\ref{dbrane}) we see that 
they are identical except for the five-sphere  which is changed
 to $T^{(1,1)}$\footnote{They also have the same $C_{0123}$  component
 but not the same components of
$C_{4}$ in the transverse directions.}. If we now deform this brane
solution electrically (magnetically) and then take the NCOS (NCYM)
limit, we will obtain the same open string metric, coupling constant
and $\tilde{\Theta}$ as in the previous subsection. Because we start with a
solution which gives identically vanishing potential for a probe
brane, we therefore know that the
{$\scr N$}=1 NCOS and NCYM theories have to be S-dual
\footnote{See Section 5 for a discussion about exact S-duality.}.

We see here that starting with a D3-brane which has a near-horizon
geometry which is a Freund--Rubin compactification of type IIB
supergravity is relatively easy to deform and investigate, because of
its similarity with the maximally supersymmetric D3-brane. 

\subsection{Electric deformation of D$1$--D$5$ system}

The D$1$--D$5$ system (with lorentzian signature) is not possible to
deform magnetically in order to obtain a 1+1 dimensional NCYM
theory. This is obvious since turning on a B-field in the D-string
directions gives an electric deformation. To obtain a NCYM theory one
has to start with a D$1$--D$5$ brane configuration with euclidean
signature. This was done in \cite{maldaR}.

To give the electrically deformed D$1$--D$5$ solution we start with
the undeformed D$1$--D$5$ solution:

\bea ds^2&=&(H_1 H_5)^{-{1\over 2}}dx^2+H_1^{1\over 2}H_5^{-{1\over
2}}dy^2+(H_1 H_5)^{1\over 2}(dr^2+r^2d\Omega_3^2)\ ,\nonumber\\
e^{2\phi}&=&g^2{H_1\over H_5}\ ,\qquad H_{1,5}=1+{\alpha'g
R_{1,5}^2\over r^2}\ ,\nonumber\\ C_2&=&{1\over g H_1}dx^0\wedge
dx^1+2\alpha'R_5^2\epsilon_2\ ,\\ C_6&=&\left[{1\over g H_5}
dx^0\wedge dx^1 + 2\alpha'R_1^2\epsilon_2\right]\wedge dy^1\wedge
dy^2\wedge dy^3\wedge dy^4\ , \nonumber \eea

where $d\Omega_3^2$ and $d\epsilon_2$ are the metric and the volume
element on the transverse $S^3$. Taking $\theta^{\mu\nu}$ to be
non-zero in the $0,1$-directions in (\ref{r21})-(\ref{tildec})
yields the following electrically deformed solution:

\bea \label{D1-D5} d\tilde{s}^2&=&(H_1H_5)^{-1/2}\frac{1}{h} dx^2+
H_1^{1/2}H_5^{-1/2}dy^2+(H_1H_5)^{1/2}
(dr^{2}+r^{2}d\Omega^{2}_{3}),\nonumber\\
e^{2\tilde{\phi}}&=&g^{2}\frac{H_1}{H_5 h}\ ,\qquad
h=1-\left(\theta\over\alpha'\right)^2(H_1H_5)^{-1}\ ,\nonumber\\
\tilde{B}_2&=&{\theta\over  \alpha'H_1 H_5 h} dx^0\wedge dx^1\
,\nonumber\\ \tilde{C}_0&=&-{\theta\over \alpha' g H_1}\ ,\\
\tilde{C}_2&=&{1\over g  H_1 h}dx^0\wedge dx^1+ 2\alpha'
R^2_5\epsilon_{2}\ ,\nonumber\\
\tilde{C}_4&=&-{\theta\over\alpha'}\left(\frac{1}{g H_5 } dy^1\wedge
dy^2\wedge dy^3 \wedge dy^4+ {2\alpha'R^2_5\over H_1H_5 h} dx^0\wedge
dx^1 \wedge \epsilon_2\right)\ , \nonumber\\
\tilde{C}_6&=&\left(\frac1{gH_5h}dx^0\wedge
dx^1+2\alpha'R_1^2\epsilon_2\right) \wedge dy^1\wedge dy^2\wedge
dy^3\wedge dy^4\ ,\nonumber\\ \tilde{C}_8&=&-{\theta \over
\alpha'gH_1H_5h} dx^0\wedge dx^1\wedge dy_1\wedge dy_2\wedge
dy_3\wedge dy_4\wedge \epsilon_2\ .\nonumber \eea

The near-horizon region of this solution is obtained by keeping the
following quantities fixed as $\alpha' \rightarrow 0$:

\bea g\ ,\quad \tilde{x}^{\mu}={x^{\mu}\over \sqrt{\alpha'}}\ ,\quad
\tilde{y}^k={y^k\over \sqrt{\alpha'}}\ ,\quad
\tilde{r}={r\over\sqrt{\alpha'}}\ , \eea

and take the critical scaling limit

\beq \frac{\theta}{\alpha'} \rightarrow 1\ , \eeq

such that

\beq h=1-(H_1H_5)^{-1}\ .\eeq

The metric (\ref{D1-D5}) interpolates between $AdS_{3}\times
S^{3}\times T^{4}$ in the IR ($\tilde{r}\sim 0$) and an array of
F$1$-strings in the UV ($\tilde{r}\sim \infty$), which are stretched
in the $x^1$ direction and wrapped on $T^4$. In the UV limit
$\tilde{r}\rightarrow\infty$ we recognize (\ref{ncos}) by
setting $\epsilon=h$ and $\lambda=1$. From this solution we calculate
the open string data in (\ref{osd1}) and (\ref{osd2}):

\beq \label{metric1}
\frac{ds^2(\tilde{G})}{\alpha'}=(H_1H_5)^{-1/2}\eta_{\mu\nu}d\tilde{x}^\mu
d\tilde{x}^\nu\ ,\eeq \beq\tilde{G}_{\rm O}=g\sqrt{\frac{H_1}{H_5}}\
,\qquad \tilde{\Theta}=\epsilon^{\mu\nu} 
\partial_{\mu}\partial^{'}_{\nu}\ .\label{metric2}\eeq

The open string metric and coupling indeed approaches constant values
when $\tilde{r}\rightarrow\infty$, as expected for a NCOS theory, and
the noncommutativity parameter remains fixed along the flow to the
IR. We also note that by introducing a fundamental (fixed) constant
with unit length, and using this to define fixed canonical coordinates
it follows from (\ref{metric1}) and (\ref{metric2}), that the Regge
slope and noncommutativity parameter of the NCOS are equal.

\section{Deformation of near-horizon geometries\label{sec:nhdef}}

In many situations one only has access to the brane solutions in the
(undeformed) near-horizon region, such as for example the various
$AdS_5$ vacua of type IIB supergravity. Since these types of solutions
are dual to field theories, and since field theories do not
admit spatio-temporal noncommutativity deformations
\cite{rabinovici}, the deformation procedure described above is only
directly applicable in the magnetic case, whereas an electric
deformation will lead to a singularity at a finite scale. This may be
understood from the form of the line element in the brane directions
(\ref{vierbein}), that is unbounded from above in the near-horizon
region, \ie $0<{\cal E}<\infty$. Hence, for any electric deformation with
finite parameter $\theta/{\alpha'}$---no matter how small---there will be a
critical scale $E_{\rm crit}=\alpha'/\theta$ where  the deformed
configuration has an essential singularity (note that in fixed units
the critical scale is $u_{\rm crit}=1/\sqrt{\theta}$).

Interestingly,  an electric deformation of the $AdS_5\times S^5$ with
parameter $\theta/\alpha'$ is S-dual to the near-horizon limit of a
magnetically deformed D$3$-brane metric with harmonic function with
negative integration constant $-\left({\theta\over
\alpha'}\right)^2$. The metrics yield (\ref{ncym}) and (\ref{ncos}),
but are pathological in other ways: the asymptotic electric geometry
is singular instead of the smeared string (\ref{smeared}) and the
magnetic probe brane has infinite negative vacuum energy in the UV so
it is unstable.

In the case of D$3$-branes, we now wish to find some criteria for when
the strong coupling limit of a noncommutative field theory on the
brane has a weak coupling dual which is a noncommutative open string
theory with an electric supergravity dual that is related
by type IIB S-duality to the magnetic  supergravity dual. This is 
true in the maximally supersymmetric case (provided the axion is rational
\cite{luroy}). As we shall show next,
under certain conditions the same is true also for extremal D$3$ branes
with less symmetry. In order to examine this in detail, we start
from the general D$3$-brane configuration in (\ref{bc1}) 
and assume (\ref{extr}). We make one further main 
assumption, namely that the axion is
vanishing (which actually could be relaxed to a rational axion \cite{luroy}):

\beq C_0=0\ .\eeq

We then find (see Appendix \ref{app:3} for details)  that the magnetic
deformation of the brane and the electric deformation of its
S-dual, are S-dual (modulo a diffeomorphism)  provided that the
undeformed dilaton is constant and the zero-force
condition is obeyed:

\beq e^{\phi}=g={\rm constant}\ ,\quad V=g^{-1}\sqrt{-\det
g_{\mu\nu}}-\omega=0\ . \label{zfc} \eeq

We also learn that the magnetic deformation parameter $\theta$ and the
electric parameter $\tilde{\theta}$ are related as follows:

\beq \tilde{\theta}=\frac{\alpha'g}{\sqrt{1+\left({\alpha'\over
\theta}\right)^2}}\equiv \sin\nu~ \alpha' g\ .\label{tilth} \eeq

We also find certain conditions on the metric and the transverse
potentials in the undeformed
configuration (see Eqs. (\ref{tfe}), (\ref{le1}) and (\ref{le2}) in Appendix
\ref{app:3}), and that the coordinates $\tilde{x}^\mu$ in the electric
supergravity dual and $x^\mu$ in the magnetic dual must be related as
follows:

\beq \tilde{x}^\mu = \sqrt{\cos\nu} x^\mu\ ,\quad 
\tilde{r}=\frac{r}{\sqrt{\cos\nu}}\ .\eeq

All this implies that the magnetic near-horizon limit (\ref{mnh})
is mapped under S-duality to the electric near-horizon limit
(\ref{enh}). We remark that (\ref{enh}) only requires
$\tilde{\theta}/\alpha'\rightarrow 1$, whereas the precise critical
scaling in (\ref{tilth}) is required by S-duality.  Hence for a D$3$ brane
configuration obeying (\ref{zfc}), (\ref{tfe}), (\ref{le1}) and (\ref{le2}),
the magnetic and electric near
horizon geometries are S-dual, and by exploiting this fact in the
extreme UV limit it follows that the corresponding NCYM and NCOS
theories must therefore be  S-dual. We can summarize the above results
in the following commutative diagram:

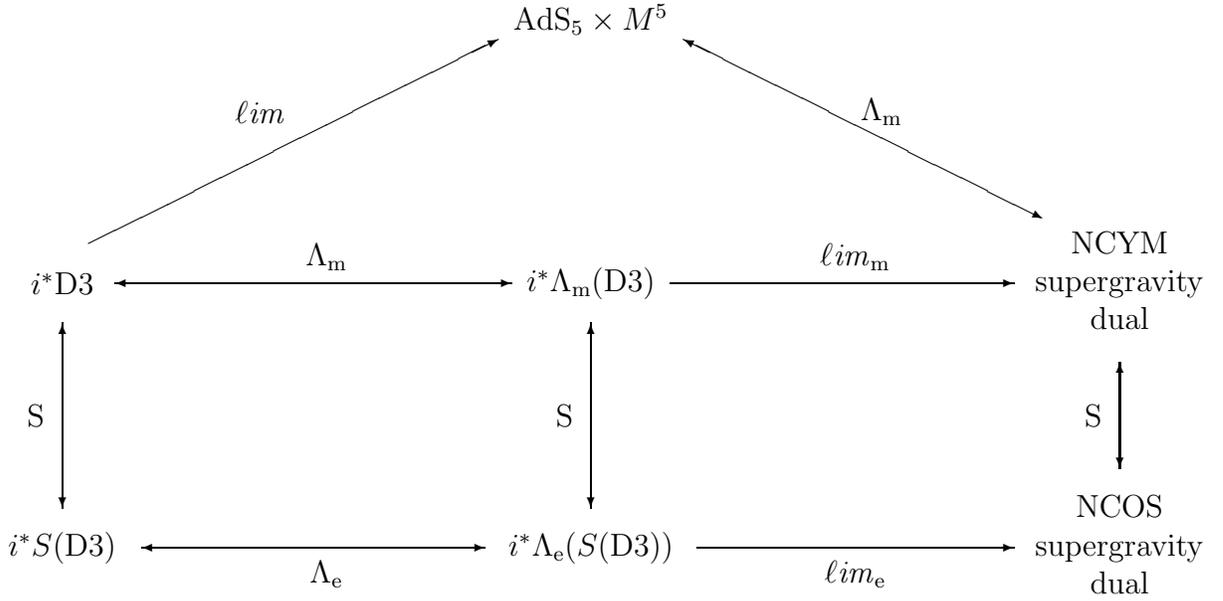
\begin{picture}(400,250)(0,-30)

\put(400,100){\makebox(0,0){$\begin{array}{c}{\rm NCYM}\\{\rm
supergravity}\\{\rm dual}\end{array}$}}

\put(400,0){\makebox(0,0){$\begin{array}{c}{\rm NCOS}\\{\rm
supergravity}\\{\rm dual}\end{array}$}}

\put(200,100){\makebox(0,0){$i^*\Lambda_{\rm m}({\rm D}3)$}}

\put(200,0){\makebox(0,0){$i^*\Lambda_{\rm e}(S({\rm D}3))$}}

\put(0,100){\makebox(0,0){$i^*{\rm D}3$}}

\put(0,0){\makebox(0,0){$i^*S({\rm D}3)$}}

\put(100,100){\vector(1,0){70}}\put(100,100){\vector(-1,0){80}}
\put(100,110){\makebox(0,0){$\Lambda_{\rm m}$}}

\put(100,0){\vector(1,0){60}}\put(100,0){\vector(-1,0){70}}
\put(100,-10){\makebox(0,0){$\Lambda_{\rm e}$}}

\put(230,100){\vector(1,0){130}} \put(300,110){\makebox(0,0){$\ell{\it
im}_{\rm m}$}}

\put(240,0){\vector(1,0){120}} \put(300,-10){\makebox(0,0){$\ell{\it
im}_{\rm e}$}}

\put(0,50){\vector(0,-1){35}}\put(0,50){\vector(0,1){35}}
\put(-10,50){\makebox(0,0){S}}

\put(200,50){\vector(0,-1){35}}\put(200,50){\vector(0,1){35}}
\put(190,50){\makebox(0,0){S}}

\put(400,50){\vector(0,-1){20}}\put(400,50){\vector(0,1){20}}
\put(390,50){\makebox(0,0){S}}

\put(10,115){\vector(2,1){155}} \put(75,165){\makebox(0,0){$\ell{\it
im}$}}

\put(370,125){\vector(-2,1){135}}\put(370,125){\vector(2,-1){1}}
\put(310,165){\makebox(0,0){$\Lambda_{\rm m}$}}

\put(200,200){\makebox(0,0){AdS$_5\times M^5$}}

\end{picture}

Here $\Lambda_{\rm m}$ and $\Lambda_{\rm e}$ denote the magnetic and
electric deformations, and $\ell{\it
im}_{\rm m}$ and $\ell{\it im}_{\rm e}$ the magnetic and electric near
horizon limits (\ref{mnh}) and (\ref{enh}), respectively. The field
theory near-horizon limit $\ell{\it im}$ is identical to the magnetic
near-horizon limit only for $\theta=0$. The operation $i^*$, where
$i$ is defined by (\ref{locemb}), indicates 
that we are only considering
the brane configuration at the vanishing locus of the brane potential
in (\ref{zfc}). Importantly, from the discussion in Section
\ref{sec:zfc} (see eq. (\ref{tilv})) it follows that $i^*$ commutes
with the deformations $\Lambda_{\rm m}$ and $\Lambda_{\rm e}$, so that
this restriction is indeed well-defined, in the sense that the NCYM
and NCOS supergravity duals have well-defined UV limits.

As already mentioned, the simplest non-trivial example of the 
above S-duality is the maximally symmetric extremal D$3$-brane. 
We shall next turn to a non-trivial example involving extremal 
branes with less symmetry and a non-trivial probe brane potential.

\section{Open strings in a ${\cal N}=1$ background\label{sec:n=1}}

In this section we examine noncommutative deformations of 
the Pilch--Warner (PW) solution
\cite{pilch}. The PW solution has conformal ${\cal N}=1$ supersymmetry and is
a warped solution with a four-dimensional vanishing locus in the internal
space.

\subsection{The PW solution\label{sec:pw}}

We use the conventions of \cite{peet}, except that we use $(-+\cdots
+)$ signature. The undeformed supergravity dual of the
${\scr N}=1$ SCFT \cite{pilch} in the string frame is given by:

\begin{eqnarray}\label{warp2}
ds^{2}&=&\alpha'\Omega^2\left[\frac{u^2}{R^2}
\eta_{\mu\nu}dx^{\mu}dx^{\nu}+\frac{R^2}{u^2}du^2+ R^2 d
\hat{s}^2_5\right],\nonumber\\ e^{2\phi}&=&g^2\ ,\quad
B_2=\alpha'\beta_2\ ,\quad C_2=\alpha'\gamma_{2},\nonumber\\
C_4&=&(\alpha')^{2}\left[\frac{k}{g}\frac{u^4}{ R^{4}} dx^0\wedge
dx^1\wedge dx^2\wedge dx^3+\gamma_4\right],\\
C_6&=&(\alpha')^3\left[\alpha_2\wedge dx^0\wedge dx^1\wedge dx^2\wedge
dx^3\right]\nonumber \, ,
\end{eqnarray}

where $ds_5^2$ is the metric on a deformed five-sphere and
$\Omega^{2}$ is a `warp-factor' depending on one of the sphere
coordinates:

\bea d\hat{s}_5^2&=&{2\over
3}\left\{\left(d\alpha^2+{\cos^2\alpha\over
3-\cos{(2\alpha)}}((\sigma^1)^2+(\sigma^2)^2) +
{\sin^2{(2\alpha)}\over
(3-\cos{(2\alpha)})^2}(\sigma^3)^2\right)\right.\nonumber\\
&&+\frac23\left.\left(d\phi+{2\cos^2\alpha\over
3-\cos{(2\alpha)}}\sigma^3\right)^2\right\}\ ,\label{intmetric}\eea
\beq \Omega^2=2^{1\over 3}\sqrt{1-\frac13 \cos{(2\alpha)}}\ .  \eeq

Here $0\leq \alpha \leq \frac{\pi}2$ and $\sigma^i$, $i=1,2,3$ are the
$SU(2)$ invariant forms satisfying
$d\sigma^1=\sigma^2\wedge\sigma^3$. The constants\footnote{There is a
typographical error in the expression for the five-form normalization
constant $m$ given in \cite{pilch}; the correct value is
$m=2^{\frac{10}3}3^{-2}R_0^{-1}$ which corresponds to the value of $k$
given below.} are

\beq R=2^{-\frac53}3R_{0}\ ,\quad R_{0}^{4}=4\pi g_{s}N\ ,\quad
k=2^{5\over 3}3^{-1}\ , \eeq

where $R_0$ is the radius of the round five-sphere in the ${\cal
N}=8$ vacuum. Note that all quantities have been given in units of
$\alpha'$ such that they are fixed in the near-horizon region. Let us
therefore define

\beq U={\Omega u\over R}\ ,\quad \omega ={ku^4\over g R^4}\ . \eeq

The zero-force condition (\ref{zfc}) implies

\beq V={u^4\over gR^4}(\Omega^4-k)=0\ \Rightarrow \alpha=0\
. \label{pwpot}\eeq

From (\ref{intmetric}) it follows that the vanishing locus is a
codimension $2$ manifold in the transverse space, with frame $du$,
$\sigma^{1,2}$ and $d\phi+\sigma^3$.

\subsection{Deforming the PW solution}

We now carry out the transformation procedure described in Section
\ref{sec:nondef} to determine the magnetically deformed near-horizon
geometry

\bea \label{warp3}
ds'^{2}&=&\alpha'\Omega^2\left[\frac{u^2}{R^2}\left(-(dx^0)^2+(dx^1)^2+
\frac{1}{h}((dx^2)^2+(dx^3)^2)\right)\right.\nonumber\\
&&\left.\qquad+\frac{R^2}{u^2}du^2+R^2
d\hat{s}_5^2\right]\ ,\nonumber\\ 
e^{2\phi'}&=&\frac{g^2}{h}\ ,\nonumber\\ 
B'_2&=&\alpha'\left[-\frac{\theta U^4}{ h}dx^2\wedge
dx^3+\beta_2 \right]\ ,\quad C'_2=\alpha'\left[-\theta\omega
dx^0\wedge dx^1+\gamma_2\right]\ ,\\
C'_4&=&(\alpha')^2\left[\frac{\omega}{h}dx^0\wedge dx^1\wedge
dx^2\wedge dx^3+\gamma_{4}+\frac{\theta U^4}{h}\gamma_{2}\wedge
dx^{2}\wedge dx^{3}\right.\nonumber\\
&&\left.\qquad-\theta\alpha_2\wedge dx^{0}\wedge dx^{1}\right]\
,\nonumber\eea

where the deformation parameter $\theta$ has dimension (length)$^2$
and we have defined

\beq h=1+\theta^2 U^4\ .\eeq

The solution (\ref{warp3}) on the vanishing locus (\ref{pwpot}), 
is interpreted as the supergravity dual of an 
${\cal N}=1$ noncommutative Yang-Mills theory.  

As discussed in Section \ref{sec:nhdef}, we can now obtain the
supergravity dual of an ${\scr N}=1$ electrically deformed theory
by S-dualizing the solution (\ref{warp3}). The S-duality rules are
given by:

\begin{eqnarray}\label{Srul}
d\tilde{s}^2&=&e^{-\phi'}ds'^2\ ,\nonumber\\
e^{\tilde{\phi}}&=&\frac{e^{\phi'}}{C'^{2}_0+e^{2\phi'}}\ ,\quad
\tilde{C}_0=-\frac{C'_0}{C'^{2}_0+e^{2\phi'}} ,\\ \tilde{B}_2&=&C'_2\
,\quad \tilde{C}_2=-B'_2\ ,\nonumber\\ \tilde{C}_4&=&C'_4+B'_2\wedge
C'_2\ .\nonumber
\end{eqnarray}

We find the following electric near-horizon geometry:

\bea \label{warps} d\tilde{s}^{2}&=&\alpha'
\left[\tilde{\cal E}^2\left({-(d\tilde{x}^0)^2+(d\tilde{x}^1)^2\over
\tilde{h}}+ (d\tilde{x}^2)^2+(d\tilde{x}^3)^2)\right)
+{\Omega^4\over
k}\tilde{\cal E}^{-2}(d\tilde{r}^2+\tilde{r}^2d\hat{s}_5^2)
\right],\nonumber\\ e^{2\tilde{\phi}}&=&{\tilde{g}^2\over \tilde{h}}\
,\nonumber\\
\tilde{B}_2&=&\alpha'\left[-{k\over\Omega^4}\frac{\tilde{{\cal E}}^4}{\tilde{h}}
d\tilde{x}^0\wedge d\tilde{x}^1+\gamma_2\right]\ ,\quad\nonumber\\
\tilde{C}_2&=&\alpha'\left[{\Omega^4\over k}
\tilde{\omega}d\tilde{x}^2\wedge d\tilde{x}^3 -\beta_2\right]\
,\\ \tilde{C}_4&=&(\alpha')^{2}\left[ {\tilde{\omega}\over
\tilde{h}}d\tilde{x}^0\wedge d\tilde{x}^1\wedge d\tilde{x}^2\wedge
d\tilde{x}^3+\gamma_4+\beta_2\wedge \gamma_2\nonumber\right.\\
&&-\left. {k\over\Omega^4}{\tilde{{\cal E}}^4\over
\tilde{h}}d\tilde{x}^0\wedge
d\tilde{x}^1\wedge(\alpha_2+\beta_2)\right]\ ,  \nonumber \eea

where we have performed the reparametrization

\beq \tilde{x}^\mu=\theta^{-\frac12}x^\mu\ ,\quad
\tilde{r}=k^{\frac12}\theta^{\frac12}u\ , \eeq

and defined

\beq \tilde{g}=g^{-1}\ , \quad \tilde{R}=g^{-{1\over 2}}R\ ,\eeq

\beq
\tilde{\cal E}^2=\left(1+{k^2\tilde{R}^4\over\Omega^4\tilde{r}^4}\right)^{-\frac12}\
, \quad \tilde{\omega}={k\over\Omega^4}{\tilde{\cal E}^4\over \tilde{g}}\
,\quad \tilde{h}=1-\tilde{\cal E}^4\ .\eeq

This is interpreted as providing the ${\cal N}=1$ supergravity dual
on the vanishing locus (\ref{pwpot}), from which one might extract a
noncommutative open string theory with ${\cal N}=1$ supersymmetry.
To investigate this and illustrate the importance of the zero-force
condition, we proceed by computing open string quantities for a probe
brane in this background. This yields the following open string metric
($\alpha, \beta =0,1$ and $a,b=2,3$):

\beq \label{metricN1} \frac{\tilde{G}_{\mu \nu}}{\alpha'}=\tilde{\cal E}^2
\left[\left(
1+\frac{\tilde{\cal E}^4}{\tilde{h}}(1-{k^2\over\Omega^8})\right)
\eta_{\alpha \beta} \oplus \delta_{ab} \right]\ ,
\label{pwosm}\eeq

open string coupling:

\beq \label{coupN1} \tilde{G}_{\rm O}=\tilde{g}\sqrt{
\left|1+\frac{\tilde{\cal E}^4}{\tilde{h}}(1-{k^2\over\Omega^8})
\right|}\ , \eeq

and the spatio-temporal noncommutativity parameter

\beq \tilde{\Theta}^{\alpha\beta} =
\epsilon^{\alpha\beta}{k\over\Omega^4}\left(1+
\frac{\tilde{\cal E}^4}{\tilde{h}}(1-{k^2\over\Omega^8})\right)^{-1}\ .
\label{pwth}\eeq

Note that (\ref{pwosm}) and (\ref{pwth}) are dimensionless in the
dimensionless $\tilde{x}^\mu$ coordinates. To have a noncommutative
open string theory one requires that these quantities are finite in
the limit $\tilde{h}\rightarrow 0$. This can only occur when

\beq \Omega^4 =k \ , \label{locus} \eeq

\ie when the zero-force condition (\ref{pwpot}) is satisfied.  When
this equation is obeyed, the open string data becomes:

\beq \frac{\tilde{G}_{\mu \nu}}{\alpha'}=\tilde{\cal E}^2 \eta_{\mu \nu} \
, \quad  \tilde{G}_{\rm O} = \tilde{g} \, , \quad
\tilde{\Theta}^{\alpha\beta} = \epsilon^{\alpha\beta} \ , \label{locusdata} \eeq

which is exactly what we should expect recalling the discussion in 
Subsection 3.2. Note that, if one starts from a full brane solution,
as in \eg \cite{Pope}, we expect to find (\ref{locusdata}) in all directions,
not only on the locus. The reason we get (\ref{pwosm}) to (\ref{pwth}) is that we here perform
 the  S-duality transformation also off the locus where it is actually not
valid, see Section 5 and Appendix B.

\section{Discussion}

In this paper we have described how the limits on a D-brane required
for a  noncommutative theory naturally arise when considering a probe
brane in an appropriate background supergravity solution. We describe
how to construct such solutions and in particular we use this construction
 to find 
deformed ${\scr N}=1$ supersymmetric solutions. It should be
stressed that the decoupling of the probe brane from the bulk only
occurs when it is in the near horizon region, with the
NC theories appearing in the UV limit of this region. Nevertheless, one
may in the near horizon region 
investigate the D-brane in terms of the open strings ending on it, and the
corresponding 
{\it{open string data}}, at any radial location. The noncommutative
deformation is shown to be independent of the probe brane radial
distance.  When one considers electrically deformed solutions one sees
that the noncommutative open string limit only occurs on D-branes that
are embedded in such a way that they have vanishing potential energy
and so obey the zero force condition. This is shown also for the
(conformal) ${\scr N}=1$ supersymmetric theory. In field theory,
the interpretation would be that the ultraviolet completion of the
theory to a noncommutative open string theory only holds for a
submanifold in moduli space where the potential energy vanishes. It
should also be noted that the primary example we have investigated,
the Pilch--Warner solution, is the dual of a very specific
${\scr N}=1$ theory that can be viewed as a relevant deformation
from the ${\scr N}=4$ theory. It would be interesting to look at
other examples of ${\scr N}=1$ theories and the results of their
deformations.

In the context of ${\scr N}=1$ theories, for which the full brane solution
is not known, only the magnetically deformed theory can be obtained by means of
the T-duality transformation applied to the near horizon solution.
 Hence one has to rely on other methods to find
the electrically deformed solution and 
in this paper this is done by applying an S-duality transformation to the
magnetic solution. However, as explained in Section 5 this works only 
on the vanishing locus of the field theory potential, and
 it therefore also follows that the NCYM and NCOS theories are S-dual 
in the UV limit only on the locus of vanishing potential
\footnote{Note that although the zero-force condition is not satisfied off 
the locus, the open string quantities defined in Section 2 nevertheless 
seem to behave the same way in all directions as follows from \eg
\cite{Pope}.}.

Throughout this paper we have always implicitly been in the large N
limit. An  interesting further area of study would be to extend the
above analysis to  include $1 \over N$ corrections and see the role
they have in the existence of the open string limit and in the
independence of the deformation on the radial direction. The
construction of the deformed solution corresponding to $1 \over N$
corrections would proceed as above beginning from an undeformed
solution to supergravity with  $\alpha'$ corrections.

Another aspect not explored in this paper is the possibility of an
SL(2,$\Z$)-covariant analysis. For example, if one considers describing
the D3-brane with open D-strings then one sees how the noncommutative
`magnetic' theory has a description in terms of noncommutative open
D-string theory at strong coupling. An SL(2,$\Z$)-invariant description
would require describing the brane fluctuations with an SL(2,$\Z$)-inert
object such as an open D3 brane. This is work in progress.

\section{Acknowledgements}

D.S.B. was supported during the initial stages of this work by
Stichting voor Fundamenteel Onderzoek der Materies (FOM), while
visiting the university of Groningen and is also grateful to Chalmers
and G\"oteborg University for support and hospitality during several visits.
D.S.B. would like to acknowledge discussions with Jan de Boer, Amit
Giveon, Troels Harmark, Chris Hull, Niels Obers and Maulik Parikh.
B.E.W.N. and M.C. are grateful for discussions with Gabriele Ferretti.
P.S. was supported by FOM. This work is partly  
supported by EU contract HPRN-CT-2000-00122 and by the Swedish and Danish 
Natural Science
Research Councils.

\appendix

\section{The noncommutative Scalar Fields \label{app:2}}

In order to derive the field redefinition in the scalar sector on a
noncommutative D$p$-brane we start from a $9$-brane in a constant
background described by the Born--Infeld lagrangian

\beq S_9=-\mu_9\int_{M^{10}} d^{10}x e^{-\check{\phi}} \sqrt{
-\det\left[ \frac{\check{g}_{MN}}{\alpha'} +
\frac{\check{B}_{MN}}{\alpha'} + F_{MN} \right]}\ , \label{s9} \eeq

where $M^{10}$ is a ten-dimensional space-time with metric
$\check{g}_{MN}$, dilaton $\check{\phi}$, two-form potential
$\check{B}_{MN}$ and field strength $F_{MN}=\partial_M A_N-\partial_N
A_M$. The $(p+1)$-dimensional Born--Infeld lagrangian on
$M^{p+1}\times {\bf R}^{9-p}$ with non-vanishing two-form fluxes
only on $M^{p+1}$ is obtained by double dimensional reduction on
$M^{10}=M^{p+1}\times T^{9-p}$ followed by decompactifying the
resulting T-dual torus keeping the effective D$p$-brane tension
$\mu_p$ fixed. Thus, if we denote the indices on $M^{p+1}$ by
$\mu=0,\dots,p$ and the indices on $T^{9-p}$ by $i=p+1,\dots,9$, and
take

\beq \check{g}_{\mu i}=0\ ,\quad \check{B}_{\mu i}=\check{B}_{ij}=0\
,\label{bij} \eeq

then we can identify the metrics and the two-form potentials on the
$M^{p+1}$,

\beq g_{\mu\nu}= \check{g}_{\mu\nu}\ ,\quad
B_{\mu\nu}=\check{B}_{\mu\nu}\ ,\label{ddr1}\eeq

and relate the metric $g_{ij}$ and the scalar coordinates $\Phi^i$
(with the dimension of energy) on the T-dual torus to the
ten-dimensional cometric and internal vector field components on
$T^{9-p}$ as follows:

\beq g_{ij}=(\alpha')^2\check{g}^{ij}\ ,\quad \Phi^i= A_i\
. \label{xi} \eeq

As a result

\beq S_9~\longrightarrow\!\!\!\!\!\!\!\!\!\!\!^{^{T^{9-p}}}~~
S_p=-\mu_p\int_{M^{p+1}}d^{p+1}x e^{-\phi} \sqrt{-\det
\left[\frac{g_{\mu\nu}}{\alpha'} + \partial_\mu \Phi^i \partial_\nu
\Phi^j \frac{g_{ij}}{\alpha'}+\frac{B_{\mu\nu}}{\alpha'}+
F_{\mu\nu}\right]}\ ,\eeq

where we have identified the T-dual dilaton and the D$p$-brane
tension as follows:

\beq \mu_p e^{-\phi}=\mu_9 {\rm Vol}_{9-p}(\check{g}) e^{-\check{\phi}}\
,\quad {\rm Vol}_{9-p}(\check{g})=\int_{T^{9-p}}d^{9-p}x
\sqrt{\det\frac{\check{g}_{ij}}{\alpha'}}\ . \label{ddr2} \eeq

We next recall \cite{sw} that if one takes the $B$-fluxes
to be non-zero in the spatial directions $m=p-1,p$ (so we assume
$p>1$) and considers the following decoupling limit on the original
$9$-brane:

\beq \ell{\it im}_{\rm 9}\ :\quad
{\check{g}_{MN}\over\alpha'}\sim\left\{\begin{array}{ll}
\epsilon^{-\frac12}&\mbox{for }M,N\neq p-1,p \\
\epsilon^{\frac12}&\mbox{else}\end{array}\right.\ ,\quad
\frac{\check{B}_{mn}}{\alpha'}\sim \epsilon^0\ ,\quad
e^{\check{\phi}}\sim\epsilon^{-1}\ , \label{ell10} \eeq

then (\ref{s9}) reduces (up to total derivatives and higher derivative
terms) to the noncommutative Maxwell action in ten dimensions:

\beq S_9 ~\longrightarrow \!\!\!\!\!\!\!\!^{^{\ell{\it im}_9}}~~~ \hat{S}_9=
-\frac14 \int_{M^{10}}d^{10}x\frac1{\check{G}_{\rm
O}}\sqrt{-\det\frac{\check{G}}{\alpha'}}
(\alpha')^2\check{G}^{MN}\check{G}^{PQ}\hat{F}_{MP}\hat{F}_{NQ}\
,\label{ncm} \eeq

where $\check{G}$ and $\check{G}_{\rm O}$ are the ten-dimensional
open string metric and coupling, and the noncommutative
ten-dimensional field strength is

\beq
\hat{F}_{MN}=\partial_M\hat{A}_N-\partial_N\hat{A}_M+\hat{A}_M\star
\hat{A}_N-\hat{A}_N\star\hat{A}_M\ . \eeq

Here the noncommutative vector potential $\hat{A}_{M}$ is related to
$A_M$ by the non-local field redefinition:

\beq \hat{A}_M=A_M-\frac12 \theta^{NP}A_N(2\partial_P A_M-\partial_M
A_P)+{\cal O}( \theta^2)\ , \label{hata}\eeq

where $\theta^{MN}$ is the ten-dimensional Poisson structure. From
Eqs. (\ref{bij})-(\ref{xi}) and (\ref{ddr2}) it follows that the
ten-dimensional open string data are related to the T-dual
$(p+1)$-dimensional open string data as follows:

\beq \check{G}_{MN}= G_{\mu\nu}\oplus \check{g}_{ij}\ ,\label{ddrosd1}
\eeq \beq \check{G}_{\rm O}= {\rm
Vol}_{9-p}(\check{G})G_{\rm O}={\rm Vol}_{9-p}(\check{g})G_{\rm O}\ , 
\label{ddrosd} \eeq

\beq \theta^{MN}=\theta^{\mu\nu}\oplus 0\ ,\quad
\theta^{\mu\nu}=\left\{\begin{array}{ll}\theta^{mn}&\mbox{for
}\mu,\nu=m,n\\ 0&\mbox{else}\end{array}\right.\ .  \label{ddrosd2} \eeq

Moreover, from (\ref{ddr1}), (\ref{xi}) and (\ref{ddr2}) it follows
that (\ref{ell10}) is equivalent to the following $(p+1)$-dimensional
decoupling limit:

\bea
\ell{\it im}_p\ :\quad
&g_{\mu\nu}/\alpha'\sim\left\{\begin{array}{ll}
\epsilon^{-\frac12} & \mbox{for }\mu,\nu=0,\dots,p-2\\
\epsilon^{\frac12} & \mbox{for }\mu,\nu=p-1,p\end{array}\right.\ ,
\label{ellp}\\ 
&g_{ij}/\alpha' \sim \epsilon^{\frac12}\ ,\quad
B_{mn}/\alpha'\sim \epsilon^0\ ,\quad
 e^{\phi} \sim\epsilon^{\frac14(5-p)}\ ,\nonumber
\eea

which we identify as the NCYM limit (\ref{ncym}) on a D$p$-brane with
$B$-flux of rank two such that (again up to total derivatives and
higher derivative terms)

\beq S_p~\longrightarrow \!\!\!\!\!\!\!\!^{^{\ell{\it im}_p}}~~~\hat{S}_p\ ,
\eeq

where $\hat{S}_p$ defines the noncommutative Yang--Mills theory on the
D$p$-brane. We have thus established the 'commutative' diagram:

\beq \mbox{
\begin{picture}(150,150)(0,-30)
\put(10,100){\makebox(0,0){$S_9$}} \put(25,100){\vector(1,0){125}}
\put(70,107){\makebox(0,0){$\ell{\it im}_9$}}
\put(160,100){\makebox(0,0){$\hat{S}_9$}}
\put(10,85){\vector(0,-1){80}} \put(-4,50){\makebox(0,0){$T^{9-p}$}}
\put(145,50){\makebox(0,0){$T^{9-p}$}}
\put(10,-5){\makebox(0,0){$S_p$}} \put(25,-5){\vector(1,0){125}}
\put(70,2){\makebox(0,0){$\ell{\it im}_p$}}
\put(160,-5){\makebox(0,0){$\hat{S}_p$}}
\put(160,85){\vector(0,-1){80}}
\end{picture}
 } \nonumber \eeq

Hence, we may obtain explicitly the form of $\hat{S}_p$ and the field
redefinition of the noncommutative scalar fields on the D$p$-brane,
by double dimensional reduction of (\ref{ncm})-(\ref{hata}) using
(\ref{ddrosd1})-(\ref{ddrosd2}), that is

\beq \hat{S}_p=- \int_{M^{p+1}} d^{p+1}x \frac1{g_{\rm YM}^2}
\sqrt{-\det{\bar{G}}}\left(\frac14\bar{G}^{\mu\nu}\bar{G}^{\rho\sigma}
\hat{F}_{\mu\rho}\hat{F}_{\nu\sigma}+ \frac12 \bar{G}^{\mu\nu}
\hat{D}_\mu\hat{\Phi}^i\hat{D}_\nu\hat{\Phi}^j \bar{g}_{ij}\right)\ ,
\label{ncs} \eeq

\bea \hat{\Phi}^i&=&\Phi^i-\theta^{mn}A_m\partial_n\Phi^i + {\cal
O}(\theta^2)\ ,\label{ncs2}\\[15pt] \hat{D}_\mu \hat{\Phi}^i&=&
\partial_\mu \hat{\Phi}^i + \hat{A}_\mu\star
\hat{\Phi}^i-\hat{\Phi}^i\star \hat{A}_\mu\ ,\nonumber \eea

and $\bar{G}_{\mu\nu}$ and $g_{\rm YM}^2$ are defined as in
(\ref{barG}) and (\ref{gYM}) and $\bar{g}_{ij}$ as in (\ref{uv1}). To
be precise, we have the following scaling behaviour:

\beq
{G_{\mu\nu}\over\alpha'}\sim\epsilon^{-\frac12}\ell^{-2}\eta_{\mu\nu}\
,\quad  {g_{ij}\over \alpha'}\sim\epsilon^{\frac12}\ell^2\bar{g}_{ij}\
,\quad G_{\rm O}\sim  \epsilon^{-\frac{p-3}4}\ , \eeq

where $\ell$ is a fixed length scale such that $g^2_{\rm YM}\sim
\ell^{p-3}$.

The condition of constant background is not crucial. The important
input is the nature of the limit (\ref{ellp}), and the fact that the
$\theta$ parameter is constant. Let us start from a D$p$-brane
supergravity dual with UV limit (\ref{ellp}), and expand the
Born--Infeld lagrangian in this limit. From the results in Section
\ref{sec:nondef} we find

\bea S_p&=&-\int d^{p+1}\xi
e^{-\phi}\sqrt{-\det\left({g_{\mu\nu}\over\alpha'}+ {B_{\mu\nu}\over
\alpha'}+S_{\mu\nu}+F_{\mu\nu}+\beta_{\mu\nu}\right)}\nonumber\\ &=&
-\int d^{p+1}\xi e^{-\phi}\sqrt{-\det \left({g_{\mu\nu}\over
\alpha'}+{B_{\mu\nu}\over \alpha'}\right)}\sqrt{\det\left(1+
\left({g\over \alpha'}+{B\over
\alpha'}\right)^{-1}(S+F+\beta)\right)}\nonumber\\ &=&-\int d^{p+1}\xi
\frac1{G_{\rm O}}\sqrt{-\det {G\over \alpha'}}
\left(\sqrt{\det(1+\theta
(F+\beta))}+(\alpha')^2G^{\mu\nu}G^{\rho\sigma}\hat{F}_{\mu\rho}\hat{F}_{\nu\sigma}
\right.  \nonumber\\
&&\left.+(\alpha')^2G^{\mu\nu}\hat{D}_\mu\hat{\Phi}^i\hat{D}_\nu
\hat{\Phi}^j \bar{g}_{ij} +\cdots\right)\ , \label{dbiexp} \eea

where

\beq S_{\mu\nu}=\alpha'\partial_\mu \Phi^i\partial_\nu \Phi^j g_{ij}\
,\quad \beta_{\mu\nu}=\alpha'\partial_\mu \Phi^i\partial_\nu \Phi^j
B_{ij}\ .\eeq

In expanding the determinant we have used the fact that $\theta$ and
$F$ are fixed in the UV while $S$, $\beta$ and $\alpha' G^{-1}$ are
subleading. The leading order, which is divergent, consists of traces
of the form ${\rm tr}((\theta(F+\beta))^n)$, where we keep
$\beta$. These terms give the divergent
$\sqrt{\det(1+\theta(F+\beta))}$-term in (\ref{dbiexp}). In the first
subleading order we find vanishing traces of the form ${\rm tr}(\theta
S(\theta (F+\beta))^n)$ and ${\rm tr}(\alpha'G^{-1}(F+\beta)(\theta
(F+\beta))^n)$, where $n$ is an integer and all possible
orderings occur. The next order, which is finite, consists of traces
of the form

\beq {\rm tr}(G^{-2}F^2(\theta F)^n)\ ,\quad {\rm tr}(G^{-1}S(\theta
F)^n)\ \mbox{and}\quad {\rm tr}(S^2\theta^2(\theta F)^n)\
,\label{structures} \eeq

multiplied by traces ${\rm tr}(\theta F)^m$, where $m,n$ are integers. 
Here we can drop $\beta$, since the scaling of the structures in
(\ref{structures}) is precisely cancelled by the determinant prefactor
in (\ref{dbiexp}), so that the $\beta$ contributions vanish in the
UV. These terms give the finite kinetic terms in
(\ref{dbiexp}), modulo derivatives of $G^{-1}$, $g_{ij}$ and $\phi$
with respect to internal coordinates $\Phi^i$. The reason is that in
forming the kinetic terms one needs to rewrite the structures in
(\ref{structures}) by moving derivatives from $F$ to the scalar fields
by integrating by parts. Note, however, that in doing so, one need 
not bother about differentiating with respect to $u$, since that 
lowers the degree of divergence (recall that $\partial_\mu u$ is fixed).
Hence, in the case of maximally symmetric extremal branes discussed
in Section \ref{sec:ex1}, this implies that the UV limit gives 
the same result (\ref{ncs}) as for a flat background. In cases with
less symmetry, as for instance the $T^{(1,1)}$ and the Pilch-Warner
solutions a more careful analysis is required to obtain the full
kinetic term for the noncommutative scalars.

\section{Conditions for S-duality\label{app:3}}

We are interested in finding some conditions for which the S-dual of
a magnetically deformed solution is equal to the electric deformation
of the S-dual of the original solution modulo a {\it{brane
isomorphism}}. By a brane isomorphism we mean a diffeomorphism that
preserves the structure of the brane solution. This may be written as
follows:

\beq S(\Lambda_{\rm m}({\rm D}3))=\varphi^*(\Lambda_{\rm e}(S({\rm
D}3)))\ \label{duality} ,\eeq

where $D3$ denotes a (not necessarily self-dual) $3$-brane
configuration as in (\ref{bc1}) (note that
$\gamma_6=0$ for a D$3$-brane), $\Lambda_{\rm m,e}$ are magnetic and
electric $O(p+1,p+1)$ transformations and $\varphi$ is a brane
isomorphism, which by definition acts as a diffeomorphism on the
transverse space and a linear transformation on the brane world volume
as follows:

\beq \varphi^*dx^\mu=M^\mu{}_\nu dx^\nu\ . \label{mmunu}\eeq

In order to simplify the analysis we assume that the axion vanishes:

\beq C_0=0\ ,\quad \gamma_8=0\eeq

For clarity, we also include the NS six-form potential, which for a
D$3$-brane must have the form:

\beq B_6=dx^0\wedge\cdots \wedge dx^3\wedge \delta_2\ .\eeq

Using the $O(p+1,p+1)$ transformation rules (\ref{r21}) and (\ref{tildec})
and the S-duality transformation rules (\ref{Srul}) we find from
(\ref{duality}) that $\varphi$ has to be a symmetry of the transverse
potentials:

\beq \varphi^*(\beta_2)=\beta_2\ ,\qquad \varphi^*(\gamma)=\gamma\ . 
\label{tfe} \eeq

By examining the conditions on the brane part of the metric and the
dilaton in the asymptotically flat region we then determine

\beq M^\mu{}_\nu=\sqrt{\cos\nu}\delta^\mu_\nu\ ,\quad
\cos^2\nu=\frac1{1+\left({\theta\over \alpha'}\right)^2}\
,\label{e1}\eeq \beq \tilde{\theta} =\pm \cos\nu~\theta g\
,\label{e2}\eeq

where $\theta$ is the magnetic parameter, $\tilde{\theta}$ the
electric parameter and $g$ the asymptotic value of the dilaton in the
undeformed D$3$-brane configuration. The conditions on the metric in
the electric and magnetic brane directions and the dilaton equation
then reads:

\beq e^{-\phi}h^{\frac12}{\cal E}^2= e^{-\tilde{\phi}}{\tilde{\cal E}^2\over
\tilde{h}} \cos\nu\ ,\quad e^{-\phi}h^{-\frac12}{\cal E}^2=e^{-\tilde{\phi}}
\tilde{\cal E}^2 \cos\nu\ , \eeq \beq h\tilde{h}=e^{2(\phi-\tilde{\phi})}\
,\eeq

where

\beq \tilde{\cal E}\equiv \varphi^*{\cal E}\ ,\quad \tilde{\phi}=\varphi^*\phi\
,\eeq

\beq h=1+\left({\theta\over \alpha'}\right)^2{\cal E}^4\ ,\quad \tilde{h}=
1-\left({\tilde{\theta}\over
\alpha'}\right)^2e^{-2\tilde{\phi}}\tilde{\cal E}^4\ .\eeq

These equations are equivalent to

\beq h\tilde{h}=1\ ,\quad \phi=\tilde{\phi}\ ,\label{res1}\eeq \beq
\tilde{\cal E}^4=\frac1{1+\cos^2\nu({\cal E}^{-4}-1)}\ .\label{le1} \eeq

We now turn to the remaining conditions on the potentials. From the
electric and magnetic directions of the two-form potentials and the
brane part of the four-form potential we get the conditions

\beq -\theta\omega=\tilde{\theta}{e^{-2\tilde{\phi}}\tilde{\cal E}^4\over
\tilde{h}}\cos\nu\ ,\quad \theta{{\cal E}^4\over
h}=-\tilde{\theta}\tilde{\omega}\cos\nu\ ,\eeq \beq
\omega={\tilde{\omega}\over\tilde{h}}\cos^2\nu\ ,\eeq

where $\tilde{\omega}=\varphi^*\omega$. Using (\ref{res1}) and
(\ref{le1}) these equations are equivalent to (\ref{zfc}). In the
six-form sector we find the condition

\beq -\theta\omega\alpha_2=\tilde{\theta}\varphi^*\delta_2\ , \eeq

which is identically satisfied when the lower rank form equations are
obeyed. Finally, the metric equation in the transverse directions
amounts to

\beq \varphi^*i^*({{\cal E}}^2 ds_6^2)= \frac1{\cos\nu} i^*({{\cal E}}^2ds_6^2))\
,\label{le2}\eeq

The condition (\ref{le2}) can be solved by taking

\beq i^*ds_6=(i^*{{\cal E}})^{-2}d\bar{s}^2_{6-n}\ ,\quad
d\bar{s}_{6-n}^2=dr^2+r^2d\bar{s}_{5-n}^2\ ,\label{locm}\eeq

where $n$ is the codimension of the vanishing locus; $r$ is a radial
coordinate acted on by the brane isomorphism by the scale
transformation

\beq \varphi^*r=\frac1{\sqrt{\cos\nu}} r\ ;\label{ee1}\eeq

and $d\bar{s}_{5-n}^2$ is an invariant metric on the remaining
$5-n$ dimensions in the locus. Then (\ref{le1}) implies that the
vierbein ${\cal E}$ is 'harmonic' on the locus:

\beq i^* {{\cal E}}^2=(1+{k(\alpha')^2\over r^4})^{-\frac12}\ .\label{harm}\eeq

In the magnetic near-horizon limit (\ref{mnh}) we can thus identify

\beq u=\frac{{{\cal E}}}{\sqrt{\alpha'}}\sim\frac{r}{\sqrt{k}\alpha'}\ .\eeq

From (\ref{e1}), which is equivalent to

\beq \varphi^*dx^\mu = \sqrt{\cos\nu}~dx^\mu\ ,\label{ee2} \eeq

and (\ref{e2}) and (\ref{ee1}) it follows that the S-dual of the
magnetic near-horizon limit is the electric near-horizon limit
(\ref{enh}).

\end{document}